\documentclass{ws-procs9x6-cpt22}
\begin{document}

\newcommand{\refeq}[1]{(\ref{#1})}
\def\etal {{\it et al.}}

\title{Status of the GINGER project}

\author{Angela D. V. Di Virgilio}\address 
{INFN Sez. di Pisa, 
Largo B. Pontecorvo 3, 56127 Pisa, Italy}
\author{On behalf of GINGER Collaboration\footnote{ GINGER Collaboration:
 INFN-Pisa and University of Pisa: Andrea Basti, Nicol\`o Beverini, Giorgio Carelli, Donatella Ciampini,  Giuseppe Di Somma,  Angela D.V. Di Virgilio,  Francesco  Fuso,  Enrico Maccioni, Fabio Morsani, Paolo Marsili, and Giuseppe Terreni. 
 INFN-Napoli,  University of Naples and CNR-SPIN: Carlo Altucci , Francesco Bajardi , Salvatore Capozziello, Alberto Porzio, Raffaele Velotta, University of Salerno: Gaetano Lambiase.
INFN-LNL: Antonello Ortolan, University of Torino: Matteo Luca Ruggiero.
INGV: Thomas Braun, Gaetano De Luca, Roberto Devoti, Giuseppe Di Stefano, Daniela Famiani, Alberto Frepoli, Aladino Govoni, and Alessia Mercuri.
 Univesity of L'Aquila: Francesco Dell'Isola, Ivan Giorgio and Marco Tallini, Politecnico delle Marche: Fabrizio Dav\'i, University of Sassari: Emilio Barchiesi and Emilio Turco.
}}

\begin{abstract}
GINGER (Gyroscopes IN GEneral Relativity), based on an array of large dimension ring laser gyroscopes, is aiming at measuring in the Gran Sasso underground laboratory the Earth angular velocity with unprecedented sensitivity in order to record the general relativity effects on top of the Earth crust; 1 part $10^9$ of the Earth surface is the goal to access the signals expected by general relativity, due to deSitter and Lense-Thirring effects, this target is also valuable for Lorentz Violation in the gravity sector. GINGER is an multi-disciplinar project. Being attached to the Earth crust, it will provide useful data for geophysical investigation, and it will be one of the instruments in the recently proposed multi-components geophysical observatory of GranSasso. It is expected that the realisation of GINGER will take 18 months, and 3 years will take to reach the relative sensitivity of 1 part in $10^{11}$.
\end{abstract}
\bodymatter
\section*{}
At present large frame Ring Laser Gyroscopes (RLGs) are the most sensitive instruments to measure absolute angular rotation rates; sensitivity of the order of prad/s with long term operation and large dynamic range have been extensively demonstrated \cite{uno, due}. \\Instrumentation based on photons has reached incredibly high sensitivity and takes advantages of the well developed industry able to produce the necessary components as high quality optics and photodetectors. Light based interferometers has reached an extremely high level of sensitivity, reliability and robustness. In this perspective, the large interferometers (e.g. LIGO and VIRGO)  are a classic  example. Different interferometer topology can be used in view of physical effects and principles under investigation. For example, let us consider a closed polygonal path, defined by 4 mirrors located at the vertices of a square, with the two light beams circulating inside the cavity in  clockwise and counter clockwise directions. In this case the interference of the two counter propagating beams brings information on the non reciprocal effects. The differences between the two paths due to non reciprocity are extremely small, and the interferometer takes advantage on having two equal paths. When the frame supporting the 4 mirrors of the ring rotates, this  interferometer is sensitive to the cavity rotation rate, an effect usually called Sagnac effect, after the french physicist Geoge Sagnac who showed it approximately 100 years ago.  Accordingly, Sagnac interferometers are a specific class of interferometers, commonly used to measure inertial  rotation velocity. However, there are other non reciprocal effects related to the propagation of the two light beams that are connected to the space time structure or symmetries, and so suitable for fundamental physics investigation.\\
The   active Sagnac interferometer, usually referred to as  ``Ring Laser Gyro'' (RLG), is  the most important instrument in this family and has demonstrated record sensitivity. \\
A RLG senses the component of the angular velocity vector $\vec{\Omega}$ along the axis of the closed polygonal cavity (typically a square), defined by the area vector. The relationship between the Sagnac frequency $\omega_s$ and the angular rotation rate $\Omega$ reads: 
\begin{equation}
\omega_s =4\frac{A}{\lambda L} \Omega \cos{\theta} \ , \\
\label{uno}         
\end{equation}
where $A$ is the area enclosed by the optical path, $L$ is its perimeter, $\lambda$ is the wavelength of the light, and $\theta$ is the angle between the area vector  and  $\vec{\Omega}$. 
    GINGER is based on 3 RLGs attached to the Earth crust. Each RLG of the array measures a different projection of the Earth angular velocity vector $\vec{\Omega}$, the scale factor of the instrument being $S = 4\frac{A}{\lambda L} $. The main objective of the instrument is to reconstruct the total angular velocity vector $\vec{\Omega}$, which contains beside the kinematic term $\vec{\Omega}_\oplus$, contributions due to gravity (e.g. deSitter and Lense-Thirring effects of General Relativity), effects of  Lorentz violation (if any), and local contributions from geophysics and geodesy. The international Earth Rotation System IERS continuously monitors and publishes the kinematic term with very high accuracy, and so gravitational theories can be tested by comparing the independent measurements of RLGs and IERS. The effectiveness of GINGER for fundamental physics investigation depends on its sensitivity, which quite often is expressed as relative precision of the measure of Earth angular rotation; it can be said that 1 part $10^{9}$ is the target to be meaningful for fundamental physics, 1 part $10^{11}$ seams at present feasible\cite{PRD,Tartaglia,Capozziello}. Recently Jay Tasson of Carlton has pointed out that GINGER could provide valuable test of Lorentz violation in the framework of Standard Model Extension (SME)\cite{quattro}.

 The geometry of the optical cavity plays a fundamental role since  it affects the absolute orientation and the scale factor S, limiting the long term response of the RLG. The cavity can be monolithic (ML), i.e. a rigid slab of low expansion glass machined with very high precision with mirrors optically contacted, or hetero-lithic (HL), i.e. done attaching together several mechanical pieces ad hoc designed (keeping the instrument at stable temperature, far from disturbances and using electronic control when necessary to improve the long term stability). The first working RLGs are based on monolithic structures, and most of the small size gyroscopes have monolithic structure, but it is clear that this choice is rather demanding in term of cost and space, and poses severe limitation in its use to built an array. The hetero lithic cavity is rigid but composed of different mechanical pieces: so far our research has been dedicated to improve the HL RLG performance.  Two prototypes have been built and extensively studied: GINGERINO\cite{GING1} and GP2\cite{GP2}. Our experimental work has indicated that it is possible to build GINGER with HL structure, since so far the following has been proven: unattended continuous operation for months, typically sub-prad/s sensitivity in 1 second of measurement, large bandwidth, fast response, in principle as fast as milli-seconds, large dynamic range, and can be oriented at will.\\
At present, several high sensitivity RLGs are operative: in Germany, G\cite{G1} of the geodetic observatory of Wettzell and ROMY in the geophysical observatory of Bavaria\cite{ROMY}, GINGERINO in the underground Gran Sasso laboratory, in Italy, ER1 of the University of Canterbury,\cite{ER1} in NewZealand, and HUST-1\cite{HUST} passive gyroscope at HUST, part of the TianQin project,  at Wuhan in China. Collaboration among the different groups already exists.
Since the angular rotation is a vector at least 3 RLGs are required.
In general, to increase the sensitivity and study the
systematic, redundancy is welcome in this kind of high sensitivity apparatus.  The GINGER array is based on 3 equal RLGs,\cite{FeasStudy}
one of the 3, called RLX, is oriented at the maximum Sagnac signal to provide the absolute value of the angular velocity, the second, RLH, with area vector along the local vertical, the third one, RLO, has area vector outside the meridian plane. The combination with the RLX signal
provides the monitoring of the relative angle of RLH and RLO with the axis of rotation, important to properly subtract the Earth rotation contribution.
 
Shot noise is the most common noise source associated to interferometers and to light detection in general. It is due to the inherent fluctuations in the number of photons associated to the measurement, or, equivalently, to the phase of the interference. RLG is different compared to standard interferometer, since it compares the frequencies of two light beams and not  their phase difference.
The standard theoretical calculation\cite{chow} for the shot noise of the RLG is based on the amplitude fluctuation of the two beams due to losses. For example the shot noise contribution to the signals is $\omega_{sn}$  in function of the cavity losses, the size of the RLG and the power inside the cavity, which can be evaluated knowing the mirror transmission, for example with 5m side,  $\mu = 2 \times 10^{-6}$, transmission $T = 5 \times 10^{-7}$ and 30 nW power, the expected shot noise is 5 prad/s $Hz^{-1/2}$. Recently the analysis based on the GINGERINO data has shown evidence of limiting noise 100 times below.\cite{SNsens} The model behind this calculation considers the interference of two totally independent laser beams whose optical cavity are disjoint with each other. It does not take into account neither backscatter or laser dynamics, that unavoidably couple the two counter-propagating beams. Moreover, shot noise is phase noise, and the model states that each beam phase noise undergoes diffusion; for this reason it is considered to act as a frequency white noise in the RLG. The two above assumptions of the model are both very conservative. Work is in progress to improve the shot noise limit estimation for RLG taking into account the coupling induced by the laser.
Fig. 1 shows a pictorial view of GINGER located inside Node B of the Gran Sasso laboratory, we expect that constructions takes 18 months to have the array operative. GINGER is interdisciplinary and financed by the collaboration between INGV and INFN.
\begin{figure}
\centering
 \includegraphics[scale=0.15]{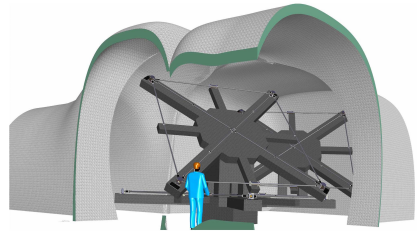}
 \caption{Pictorial view of GINGER, RLX, RLH and RLV are visible.}
 \label{GINGER}
\end{figure}

\end{document}